\documentclass{article}
\usepackage[preprint]{spconfa4}
\usepackage[utf8]{inputenc}
\usepackage{xspace}
\usepackage{xcolor}
\usepackage{stfloats}
\usepackage{graphicx}
\usepackage{amsmath}
\usepackage{pifont}
\usepackage{booktabs}
\usepackage{siunitx}
\usepackage{longtable}
\usepackage[hidelinks]{hyperref}
\usepackage{verbatim}
\usepackage[sort]{cite}
\usepackage{multirow}
\usepackage{amsmath}
\usepackage{amsfonts}
\usepackage{url}
\usepackage{flushend}

\copyrightnotice{978-1-6654-6867-1/22/\$31.00~\copyright2022 IEEE}
\name{Philipp G\"{o}tz$^{1}$, Cagdas Tuna$^{2}$, Andreas Walther$^{2}$, Emanu\"{e}l A. P. Habets$^{1}$\thanks{$^\dag$International Audio Laboratories Erlangen is a joint institution of the Friedrich-Alexander-University Erlangen-N\"{u}rnberg (FAU) and Fraunhofer IIS.
\newline Corresponding author: philipp.goetz@audiolabs-erlangen.de}}
\address{$^1$International Audio Laboratories Erlangen$^\dag$, Germany.
\\ $^2$Fraunhofer Institute for Integrated Circuits IIS, Erlangen, Germany.}

\title{AID: Open-Source Anechoic Interferer Dataset}
\author{Philipp G\"otz, Cagdas Tuna, Andreas Walther \\ Emanu\"el A. P. Habets}

\newcommand{\Hz}{\,\mathrm{Hz}\xspace}
\newcommand{\cmark}{\ding{51}}%
\newcommand{\xmark}{\ding{55}}%

\begin{document}

\maketitle

\begin{abstract}
A dataset of anechoic recordings of various sound sources encountered in domestic environments is presented. The dataset is intended to be a resource of non-stationary, environmental noise signals that, when convolved with acoustic impulse responses, can be used to simulate complex acoustic scenes. Additionally, a Python library is provided to generate random mixtures of the recordings in the dataset, which can be used as non-stationary interference signals.
\end{abstract}


\section{Introduction}
In light of the remarkable success that machine learning and deep learning has enjoyed across a broad range of audio related problems, the well-established and widespread practice of generating reverberant, noisy audio signals by adding interference to the convolution of anechoic source material and acoustic impulse responses (AIR) is as relevant as ever. This method of generating training and test data is crucially important in various fields of research such as speech enhancement \cite{benesty2011speech}, array processing \cite{luijten2019deep}, acoustic echo cancellation \cite{ivry2021nonlinear}, source localization \cite{grumiaux2021survey} and acoustic scene analysis \cite{abesser2020review,gamper2018blind,9746457}, as it offers flexible control in studying the robustness of algorithms with respect to adverse conditions. 

In the early days of acoustic signal processing, a proposed method was most often investigated using a signal of interest that is corrupted by additive, stationary noise -- for instance, produced by the receiving sensor -- at various signal-to-noise ratios (SNR). In many acoustic environments, however, the received signal is not only corrupted by stationary noise but also by non-stationary noise, which is typically introduced by a sound source located in the same acoustic environment. Such scenarios are considered in recent studies \cite{rangachari2006noise,zhang2018deep,hao2020masking}.

There are various audio datasets available that provide non-stationary noise signals. However, the vast majority of such material contains reverberation and is not suited to generate realistic, acoustic scenes from anechoic source material and AIRs. Table \ref{tab:datasets} offers a selected overview of publicly available audio data sets that contain, apart from speech, also music and environmental noise recordings. However, to the best of the authors' knowledge, the only available resource that includes anechoic recordings is the \emph{RWCP Sound Scene Database} \cite{nakamura-etal-2000-acoustical}.

In this work, we present a new dataset of anechoic recordings of a wide range of sound sources encountered in domestic environments. Complementing the dataset, we provide a Python library to generate multi-channel, random mixtures from the anechoic recordings. Additionally, modulated noise signals with a user-defined spectral slope can be generated, where the temporal envelope is computed from the anechoic noise recordings. The generated data may be used for the realistic simulation of microphone signals in challenging acoustic environments, which requires access to anechoic source material and AIRs between each sound source and microphone.

The remainder of this paper is organized as follows. In Section~\ref{sec:measurements}, the measurement process and setup are outlined, Section~\ref{sec:data_set} covers the actual dataset. The signal generator is discussed in detail in Section~\ref{sec:generator}, Section \ref{sec:conclusions} concludes this contribution.

\begin{table*}[ht]
    \centering
    \begin{tabular}{l l c c}
    \toprule
    Name & Description & Year published & Anechoic Noise \\
    \midrule
    RWCP Sound Scene Database \cite{nakamura-etal-2000-acoustical} & \begin{tabular}{@{}l@{}} Database of recordings of real-world sounds \\ and measured room impulse responses \end{tabular} & $2000$ & \cmark \\[1.1em]
    ACE Challenge \cite{ace} & \begin{tabular}{@{}l@{}} Anechoic speech, reverberant noise,\\ and AIRs recorded in seven rooms with \\ five different microphone/array types \end{tabular} & $2015$ & \xmark \\[1.6em]
    Google AudioSet \cite{gemmeke2017audio} & \begin{tabular}{@{}l@{}}Large dataset of human-labeled\\10-second sound clips\end{tabular}& $2017$ & \xmark \\[1.1em]
    ESC-50 \cite{piczak2015dataset} & \begin{tabular}{@{}l@{}} Dataset for environmental sound \\ classification 2000 5-second recordings \end{tabular}& $2017$ & \xmark \\[1.1em]
    PYRAMIC \cite{scheibler2018pyramic} & \begin{tabular}{@{}l@{}} Anechoic microphone array recordings \\ of AIRs, speech and white noise \end{tabular} & $2018$ & \xmark \\[1.1em]
    AVAD-VR \cite{thery2019anechoic} & \begin{tabular}{@{}l@{}} Anechoic recordings of different songs \\ played by a classical music ensemble \end{tabular} & $2019$ & \xmark \\[1.1em]
    \begin{tabular}{@{}l@{}} Inter-floor noise dataset \\ SNU-B36-50E \cite{choi2019classification} \end{tabular} & \begin{tabular}{@{}l@{}} Five noise sources measured \\ in a reverberant environment \end{tabular} & $2019$ & \xmark \\[1.1em]
    Neural Audio Fingerprint Dataset \cite{ahym-e477-21} & Music, AIRs and background noise & $2021$ & \xmark \\[0.5em]
    DCASE Challenge \cite{Mesaros2016_EUSIPCO} & \begin{tabular}{@{}l@{}} Several datasets for the various \\ sub-tasks that form the annual challenge \end{tabular} & $2016$ -- $2022$ & \xmark \\[0.5em]
    \bottomrule
    \end{tabular}
    \caption{Selection of publicly available audio datasets containing stationary and non-stationary noise.}
    \label{tab:datasets}
\end{table*}

\section{Measurements}
\label{sec:measurements}
The recordings of all samples in the dataset were carried out in the anechoic chamber of the Fraunhofer Institute for Digital Media Technology IDMT, Ilmenau, Germany \cite{rar}. The chamber was built in accordance with the standards DIN EN ISO 3745 and DIN EN 60268--5 using a room-in-room design resting on an elastic bearing, and has dimensions $9.4\,\mathrm{m}\times4.93\,\mathrm{m}\times6.9\,\mathrm{m}$ (L$\times$W$\times$H). The room boundary surfaces are fully lined with wedge absorbers and the room response is fully anechoic down to a frequency of $63\Hz$. A metal grid platform, used as a mounting surface for various sound sources, was installed approximately in the center of the room.

\begin{figure}[h!]
    \centering
    \includegraphics[width=\columnwidth]{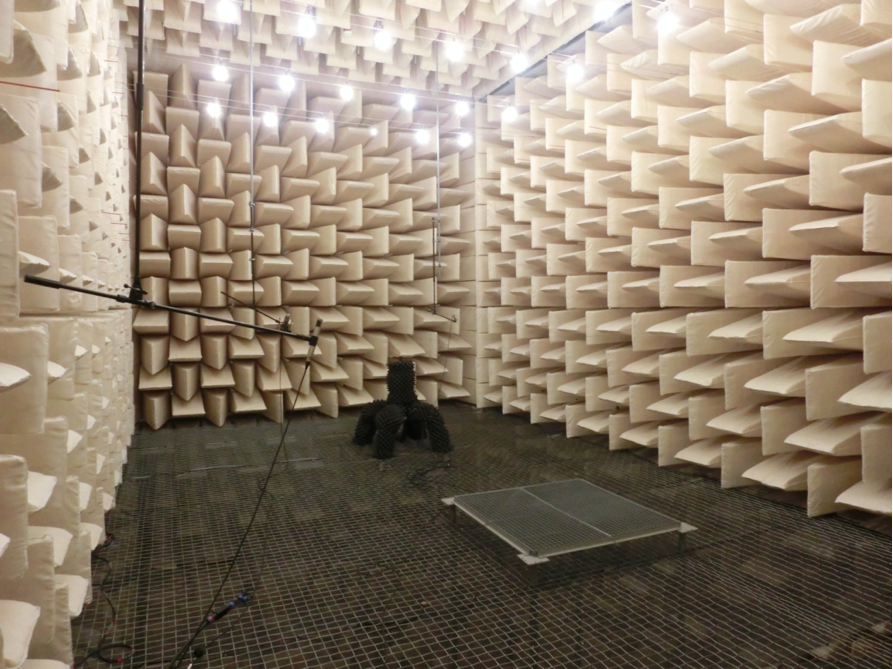}
    \caption{Anechoic chamber with walking grid and metal mounting platform, the socket of the rotating arm in the back of the room was covered in black absorbing foam to attenuate reflections.}
    \label{fig:rar}
\end{figure}

\begin{figure*}[!t]
    \centering
    \includegraphics[width=0.7\textwidth]{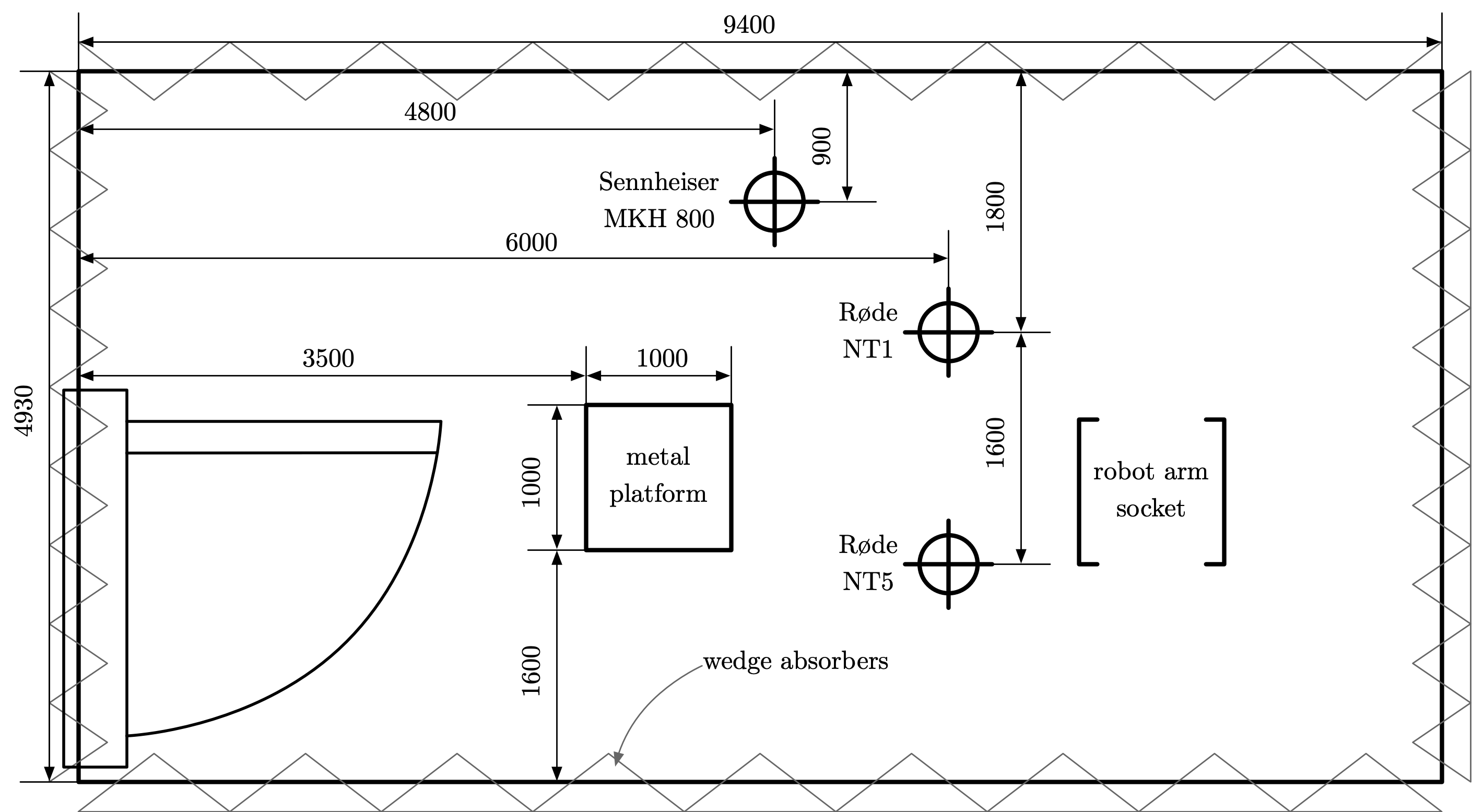}
    \caption{Top-view illustration of the recording setup in the anechoic chamber, the three microphones were positioned at approximately equal distances around the metal platform, on which all sounds were generated. The indicated distances are given in mm.}
    \label{fig:setup}
\end{figure*}

Three different microphones, an \emph{NT1} and an \emph{NT5} condenser microphone by manufacturer \emph{R{\o}de}, and an \emph{MHK 800} condenser microphone by manufacturer \emph{Sennheiser}, were used to record the various sound sources.  The \emph{Sennheiser MKH 800} offers switchable configuration of the device, in case of the recordings described herein, the directional characteristic was set to \emph{Cardioid} polar pattern, all frequency-dependent configurations, such as a low-cut at $50\,\mathrm{Hz}$ and a boost of frequencies of $8\,\mathrm{kHz}$ and above were disabled. All microphones were mounted on microphone stands that extended from the ceiling (cf. Fig.~\ref{fig:rar}), effectively attenuating acoustic coupling via non-air transmission paths. The main axes of all microphones were oriented towards the metal platform, each microphone membrane surface was oriented perpendicular to the direction of the sound sources. Figure~\ref{fig:setup} illustrates the arrangement of the metal platform and the microphones in the anechoic chamber. Using different microphones at different directions from the source, the size and diversity of the collected dataset is efficiently increased. An \emph{RME Fireface UFX II} was used to supply the phantom power to the condenser microphones and to amplify the received signals. Given the large dynamic range of the different sound sources that were recorded, the pre-amplification gains for the three microphones were individually fixed to levels such that the loudest of sounds reached peak amplitudes of approximately $-6\,\mathrm{dBFS}$. All signals were recorded at a sampling rate of $48\,\mathrm{kHz}$ and a bit-depth of $24$ bits. 

\section{Dataset}
\label{sec:data_set}
The dataset consists of anechoic recordings of $44$ different types of sound sources encountered in domestic environments, with the number of individual recordings per sound source varying between two and eleven. The sound sources, which are mostly household devices and utilities, include door keys, plastic bags, clothing, a drilling machine, an electric blender, glass jars and metal boxes but also a few human-made sounds, such as clapping, breathing, snapping or whistling. The recordings cover a wide range of timbres. Multiple sounds from every individual source were recorded by different ways of excitation, such as hitting and shaking, or switching on and off the electronic devices. 

Table \ref{tab:sound_sources} lists all sound sources and the number of different sounds contained in the dataset with an informal categorization along two perceptual attributes, transient-continuous and noisy-harmonic.

As outlined in Section~\ref{sec:measurements}, all sounds were recorded with three different microphones, resulting in $762$ individual signals with durations from $0.18\,\mathrm{s}$ up to $43\,\mathrm{s}$. The individual recordings are stored in \verb|.wav| file format and named according to \verb|soundsource_index_microphone.wav|, where \verb|soundsource| corresponds to the labels in Table~\ref{tab:sound_sources}, \verb|index| varies from $1$ to the number of individual recordings per source, indicated in Table~\ref{tab:sound_sources}, and  \verb|microphone| is either \emph{NT1}, \emph{NT5} or \emph{MKH800}.

The complete dataset, including all individual recordings, is available at \url{https://zenodo.org/record/6974033}.

\begin{table}[h]
\small
    \centering
    \begin{tabular}{l l l}
    \toprule
    & \multirow{2}{*}{\textsc{Continuous}} & \multirow{2}{*}{\textsc{Transient}} \\
    & & \\
    \midrule
    \multirow{13}{*}{\rotatebox[origin=c]{90}{\textsc{Noisy}}} & \texttt{breath[2]} & \texttt{book[11]} \\
    & \texttt{cloth[9]} & \texttt{chest\_drumming[5]} \\ 
    & \texttt{consonant[2]} & \texttt{clapping[2]} \\
    & \texttt{hairdryer[7]} & \texttt{click[1]} \\
    & \texttt{paper[12])} & \texttt{cough[2]} \\
    & \texttt{pen\_case[3]} & \texttt{foot\_steps[5]} \\
    & \texttt{plastic\_bag[2]} & \texttt{lighter[3]} \\
    & \texttt{plastic\_bottle[6]} & \texttt{lip\_flabber[1]} \\ 
    & \texttt{plastic\_wrap[5]} & \texttt{various\_tosses[10]} \\
    & \texttt{roll\_stool[6]} & \texttt{moka\_pot[13]} \\ 
    & \texttt{rubber\_band[7]} & \texttt{scissors[4]} \\
    & \texttt{vacuum\_cleaner[6]} & \texttt{finger\_snap[1]} \\
    & \texttt{zipper[4]} & \texttt{tape\_measure[4]} \\
    & & \texttt{tool\_case[3]} \\[1mm]
    \hline \\[-2ex]
    \multirow{13}{*}{\rotatebox[origin=c]{90}{\textsc{Harmonic}}} & \texttt{blender[9]} & \texttt{cutlery[11]} \\
    & \texttt{drill[10]} & \texttt{glass\_jar[10]} \\ 
    & \texttt{keys[2]} & \texttt{metal\_box[3]} \\
    & \texttt{whisk[10]} & \texttt{music\_box[3]} \\
    & & \texttt{whistle[4]} \\
    & & \texttt{plastic\_box[8]} \\
    & & \texttt{plastic\_bucket[16]} \\
    & & \texttt{plastic\_tub[4]} \\
    & & \texttt{plates[4]} \\
    & & \texttt{shoe\_squeek[2]} \\
    & & \texttt{tin\_box[16]} \\
    \bottomrule
    \end{tabular}
    \caption{Filenames of all sound sources with the number of individual recordings from each source in brackets.} \vspace{-1em}
    \label{tab:sound_sources}
\end{table}

\section{Signal generator}
\label{sec:generator}
The following section covers the motivation and implementation details of the generation of random mixtures from anechoic recordings. Furthermore, an extension of the signal generator functionality to produce modulated noise signals is discussed. A Python implementation of the random mixture and modulated noise generation is provided at \url{https://github.com/audiolabs/anechoic-noise}.

\subsection{Interference Mixture}
\begin{figure}[!t]
    \centering
    \includegraphics[width=\columnwidth]{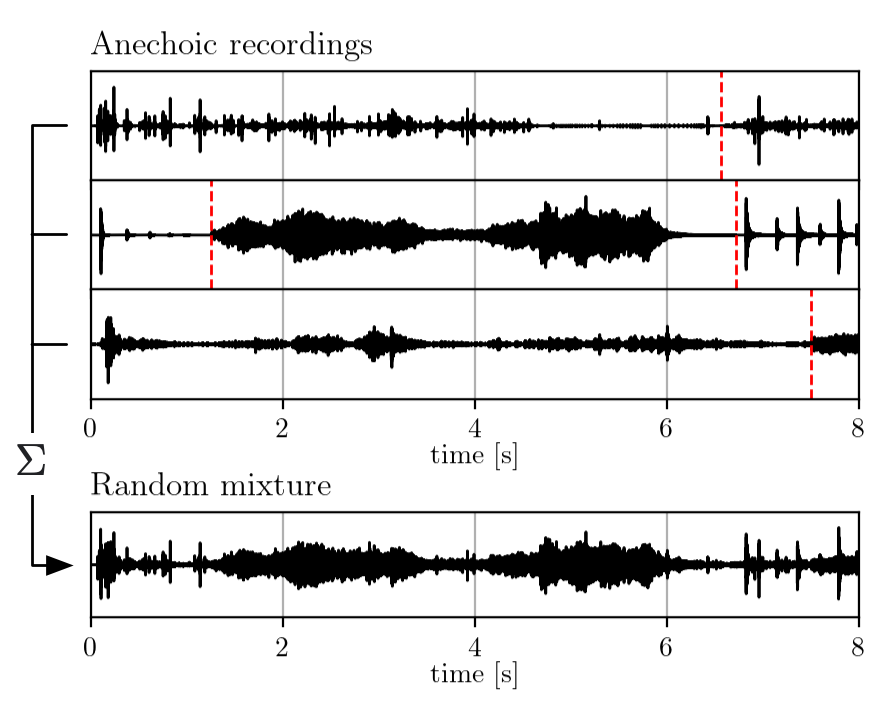}
    \caption{Arrangement of anechoic recordings that are summed into a random mixture (bottom) with $k = 3$ sounds playing concurrently, the dashed lines indicate the endpoints of individual samples.} \vspace{-1em}
    \label{fig:generator}
\end{figure}

Non-stationary interference signals used to develop and evaluate methods targeting challenging conditions are ubiquitous in many audio related fields. In many cases, the interference signal is convolved with an impulse response to simulate the sound source in a reverberant environment. However, the specific use of these signals, e.g., their representation (time vs. frequency domain), or the power ratio at which they are contrasted with the signal of interest, depends on the specific application context. To complement the presented dataset with a method to generate anechoic, non-stationary interference, the authors provide a \emph{Python} library that can be used to randomly arrange multiple anechoic noise recordings into a single channel interference signal, which is stored in \verb|wav|-file format. The number of individual recordings $k$ concurrently playing at any point in time in an interference signal can be specified by the user, providing control over the temporal density. An example of such a composition is shown in Figure~\ref{fig:generator}. In scenarios where multiple interfering sources are present, the number of single channel interference signals can be increased, with each mixture being stored in a separate \verb|wav|-file channel.

\subsection{Modulated Noise}
In most of the anechoic recordings, the signal power is distributed non-uniformly across the spectrum, resulting in considerable variation between different generated samples. However, there are scenarios in which the transient nature but not the frequency-selective characteristics of these sounds are desirable. 

To accommodate this requirement, the signal generator is extended to produce transient, colored noise signals, of which the temporal magnitude envelope is obtained from the recordings in the dataset. The envelope $e_x[n]$ of the random mixture $x[n]$ is computed via auto-recursive averaging of the signal segments of decreasing magnitude. The averaging time constant $\tau$, specified in seconds, offers control over the temporal decay rate of the noise transients:
\begin{equation}
e_x[n] = \begin{cases} 
(1 - \alpha) e_x[n-1] + \alpha \, |x[n]|,&\text{if } |x[n]| < e_x[n-1]\\
|x[n]|,& \text{otherwise,}
\end{cases}
\end{equation}
with the averaging coefficient $\alpha = 1 - \exp{(-T/\tau)}$, where $T$ denotes the sampling interval. The colored noise signal is generated from a white Gaussian noise process, $w[n]$, with a user-defined spectral magnitude slope exponent $\beta$ that exhibits a power spectral density equal to
\begin{equation}
    S(\omega_k) = \mathbb{E} \left\{\omega_k^{-\beta} \, |W(\omega_k)|^2\right\},
\end{equation}
where $W(\omega_k)$ is the discrete Fourier transform at frequency $\omega_k$ of the auto-correlation function of $w[n]$. For instance, with $\beta$ set to zero, the resulting noise signal has a white (flat) spectrum, $\beta$ set to one would result in a pink spectrum. Finally, the coloured signal is multiplied with the envelope, $e_x[n]$, to obtain the modulated noise signal.
\begin{figure}[t]
    \centering
    \includegraphics[trim={0 0.05cm 0 0},clip,width=\columnwidth]{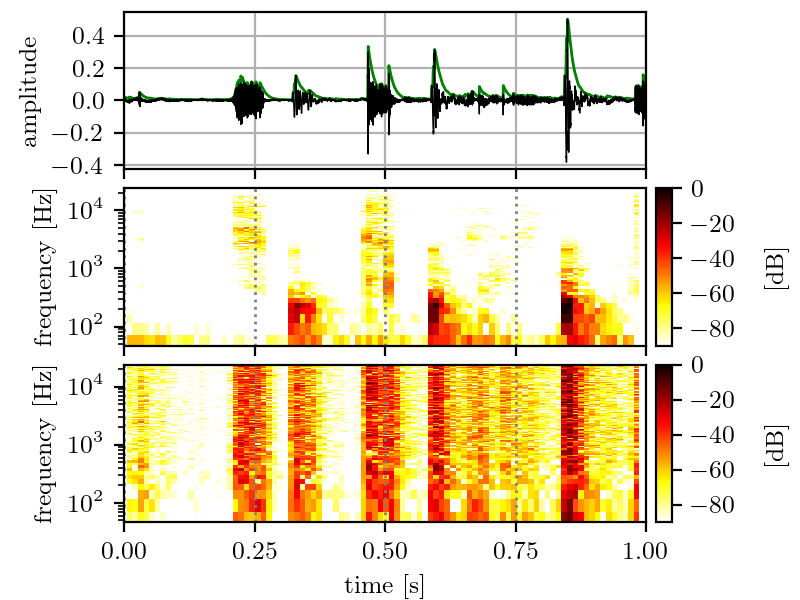}
    \caption{Top: Randomly generated non-stationary interference signal, the envelope signal with $\tau=10\,\mathrm{ms}$ shown in green; Center: Magnitude spectrogram of the random sample; Bottom: Magnitude spectrogram of the modulated white noise signal ($\beta = 0$).} \vspace{-1em}
    \label{fig:signal}
\end{figure}

\section{Conclusions}
\label{sec:conclusions}
A dataset is presented that consists of anechoic recordings of a wide variety of sound sources encountered in domestic environments. The collection of  non-stationary signals enables researchers to faithfully simulate diverse acoustic scenes by convolution of anechoic source material and acoustic impulse responses.In addition, a Python library is provided to generate random samples of anechoic sounds intended for use as non-stationary interference signals.

\bibliographystyle{IEEEbib}
\bibliography{short-strings,refs}

\end{document}